# The Nexus of Open Science and Innovation: Insights from Patent Citations

Abdelghani Maddi[*]

[*]*abdelghani.maddi@cnrs.fr*
ORCID: https://orcid.org/0000-0001-9268-8022
GEMASS – CNRS – Sorbonne Université, 59/61 rue Pouchet 75017 Paris, France.

This paper aims to analyze the extent to which inventive activity relies on open science. In other words, it investigates whether inventors utilize Open Access (OA) publications more than subscription-based ones, especially given that some inventors may lack institutional access. To achieve this, we utilized the (Marx, 2023) database, which contains citations of patents to scientific publications (Non-Patent References - NPRs). We focused on publications closely related to invention, specifically those cited solely by inventors within the body of patent texts. Our dataset was supplemented by OpenAlex data. The final sample comprised 961,104 publications cited in patents, of which 861,720 had a DOI. Results indicate that across all disciplines, OA publications are 38% more prevalent in patent citations (NPRs) than in the overall OpenAlex database. In biology and medicine, inventors use 73% and 27% more OA publications, respectively, compared to closed-access ones. Chemistry and computer science are also disciplines where OA publications are more frequently utilized in patent contexts than subscription-based ones.

## 1. Introduction

Innovation is the cornerstone of societal progress, propelling advancements in technology, economy, and scientific understanding (de las Heras-Rosas & Herrera, 2021; Jefferson et al., 2018). At the heart of this dynamic landscape lies the concept of open science, which advocates for unrestricted access to research findings and collaborative sharing of knowledge (Hecker et al., 2018). Open science represents a departure from traditional closed models of scientific dissemination, aiming to democratize access to information and accelerate the pace of discovery. Within the realm of open science, one crucial aspect deserving closer examination is the accessibility of scientific publications. Open Access (OA) to research articles has become increasingly prevalent, facilitated by initiatives such as institutional repositories, preprint servers, and open access journals. This shift towards OA has been lauded for its potential to democratize knowledge, remove barriers to information, and foster collaboration among researchers worldwide (Robinson-Garcia et al., 2020).

Despite the considerable attention paid to the benefits of open science, its specific impact on innovation remains relatively underexplored (Jahn et al., 2022). While open science initiatives strive to enhance transparency, reproducibility, and collaboration in research endeavors, the extent to which these principles translate into tangible advancements in inventive activity is less understood. This gap in understanding prompts a critical inquiry into the relationship between open access publications and inventive endeavors. Patent citations serve as a tangible indicator of the scientific knowledge base underpinning technological innovation. Known as Non-Patent References (NPRs), these citations provide insights into the sources of information and inspiration driving inventive activity. By analyzing the prevalence and utilization of open access publications within patent citations, we can gain valuable insights into the role of open science in shaping innovative processes.

In this paper, we aim to address this gap in knowledge by conducting a comprehensive analysis of patent citations across various disciplines. Leveraging data from the database curated by (Marx, 2023), supplemented by OpenAlex data, we focus on discerning patterns and trends in the citation of open access versus subscription-based publications. Specifically, we seek to answer key questions such as whether inventors exhibit a preference for open access publications, how this preference varies across disciplinary boundaries, and what implications these findings hold for the promotion of open science and innovation.

In contrast to the study conducted by (Jahn et al., 2022), which encompasses all Non-Patent References (NPRs), including those not directly related to the invention or introduced by examiners, our study focuses specifically on NPRs cited by inventors themselves and within the body of the invention text. Moreover, our analysis introduces the use of the normalized indicator NOAI (Normalized Open Access Index) to account for disciplinary disparities. This approach allows for a more nuanced understanding of the utilization patterns of open access publications within the context of patent citations, taking into consideration disciplinary differences and providing insights into the reliance on open access literature across various fields of research.

Through our investigation, we endeavor to contribute to a deeper understanding of the mechanisms driving innovation in the 21st century and the role of open science therein. By shedding light on the relationship between open access publications and inventive activity, we aim to inform policy discussions, scholarly discourse, and practical strategies aimed at fostering a more open and collaborative research ecosystem. Ultimately, our findings aspire to empower stakeholders with actionable insights to leverage open science principles for stimulating innovation and addressing societal challenges.

## 2. Data

This study relies on data provided by (Marx, 2023), which includes a vast collection of 6 million scientific publications cited in patents, focusing on what are called Non-Patent References (NPRs). This dataset is unique because it distinguishes between references cited by examiners and those cited by inventors (applicants). Additionally, it provides information on where each citation is located within the patent text (front page, body, or both), giving us valuable context.

To focus our analysis, we selected publications cited in the body of patent texts by applicants. This choice was made because citations on the front page of patents, especially in the U.S. Patent and Trademark Office (USPTO), are often added by examiners and may not directly reflect the inventors' contributions. Research has shown that citations made by inventors within the body of the patent text, which describe the invention in detail, are the most relevant to the innovation process (Bryan et al., 2020; Quemener et al., 2024).

To enrich our analysis, we used the OpenAlex API (https://docs.openalex.org/) to gather additional metadata from these publications. This included information about the specific domains (referred to as "concepts" in OpenAlex) within the highest-level aggregation, which encompasses 19 domains. OpenAlex assigns each article one or more concepts with a relevance score ranging from 0 to 1. We considered all concepts with a non-zero score. Additionally, we calculated a fractional count to measure the weight of each discipline based on the score developed by (Klebel & Ross-Hellauer, 2023).

Our final dataset comprises 961,069 publications cited in patents, with 861,729 having a DOI. This sample size ensures the reliability of our analysis and provides meaningful insights into the use of open access publications in inventive activity.

## 3. Methods

To assess the extent to which OA publications are more or less concentrated among patent citations, we employed a method based on the double ratio, as outlined in Maddi's (2020) study. This double ratio is formalized as the Normalized Open Access Index (NOAI), which compares the proportion of OA publications within each discipline among patent citations to the corresponding proportion in the entire OpenAlex database. Specifically, the NOAI is calculated by dividing the proportion of OA publications in a given discipline among patent citations by the same proportion in the entire OpenAlex database. Thus, a NOAI greater than 1 indicates an over-representation of OA publications among patent citations for that discipline, while a NOAI less than 1 suggests under-representation.

For example, if the NOAI for a particular discipline is 1.3, it signifies that OA publications are cited in patents 30% more frequently than the average for OA publications in the entire OpenAlex database. Conversely, a NOAI of 0.8 would indicate that OA publications are less frequently cited in patents within that discipline compared to the overall average for OA publications. This approach enables us to quantitatively and comparatively understand the concentration of Open Access publications within the inventive activity, highlighting differences in the representation of OA publications among patents across disciplines.

## 4. Results

This section comprises two main parts aimed at providing an understanding of the utilization patterns of Non-Patent References (NPRs) across various disciplines. Firstly, we present descriptive statistics regarding the distribution of NPRs by discipline, shedding light on the prevalence of these references within different fields of research. Secondly, we delve into the outcomes of calculating the Normalized Open Access Index (NOAI).

*4.1. descriptive statistics*

Table 1 provides a significant insight into the utilization of Non-Patent References (NPRs) across various disciplines, along with comparisons between the rates of Open Access (OA) publications within the NPRs dataset and the entire OpenAlex database.

Firstly, examining the distribution of the total number of NPR publications by discipline reveals substantial variation. Disciplines such as Biology, Chemistry, and Medicine stand out with higher numbers of NPRs, representing 26.8%, 23.0%, and 16.4% of the total NPRs, respectively. However, some disciplines, such as Art, History, Geography, and Sociology, are less represented, each accounting for less than 1% of the total. This disparity was to be expected, as NPRs are publications cited in patents, meaning they are cited based on the specific application domains of patents. Thus, disciplines less directly related to technical or industrial applications may naturally be less frequently cited in patents.

Secondly, the comparison of OA rates between the NPRs dataset and the entire OpenAlex database reveals interesting trends. While the overall OA publication rate in OpenAlex is 21.4%, the OA rate within the NPRs dataset is significantly higher, reaching 32.4%. This suggests that NPRs have a higher propensity to include OA publications compared to the entire database. However, it is important to note that this trend varies by discipline. For example, Medicine displays an OA rate of 36.1% in the NPRs dataset, while Philosophy has a much lower OA rate of 15.9%. This disparity underscores the importance of considering disciplinary specificities when analyzing the use of OA publications in NPRs. Furthermore, it's worth noting a curious result observed in Material Science and Physics where the OA rate within NPRs is lower than the OpenAlex average. These cases merit further investigation as they deviate from the general trend observed across disciplines.

**Table 1: Distribution of Non-Patent References – NPRs, by OpenAlex domains, fractional counting**

| Discipline | NPRs - Non-OA | NPRs - OA | # in NPRs dataset | % in NPRs dataset | % OA in NPRs | % OA in all OpenAlex database |
|---|---|---|---|---|---|---|
| Biology | 109 193.79 | 121 893.04 | 231 086.84 | 26.8% | 52.7% | 30.4% |
| Chemistry | 145 082.55 | 52 862.01 | 197 944.55 | 23.0% | 26.7% | 23.7% |
| Medicine | 90 432.60 | 51 085.19 | 141 517.78 | 16.4% | 36.1% | 28.4% |
| Computer science | 72 385.94 | 17 617.05 | 90 002.99 | 10.4% | 19.6% | 18.2% |
| Materials science | 66 965.56 | 12 159.08 | 79 124.64 | 9.2% | 15.4% | 21.2% |
| Physics | 24 394.83 | 6 639.40 | 31 034.24 | 3.6% | 21.4% | 24.6% |
| Mathematics | 12 183.89 | 4 308.57 | 16 492.46 | 1.9% | 26.1% | 27.2% |
| Psychology | 12 411.54 | 3 328.04 | 15 739.58 | 1.8% | 21.1% | 23.5% |
| Engineering | 11 733.82 | 1 962.80 | 13 696.62 | 1.6% | 14.3% | 21.1% |
| Political science | 5 249.49 | 998.41 | 6 247.90 | 0.7% | 16.0% | 18.1% |
| Philosophy | 4 815.92 | 911.86 | 5 727.78 | 0.7% | 15.9% | 20.5% |
| Geology | 4 417.90 | 1 026.32 | 5 444.22 | 0.6% | 18.9% | 19.6% |
| Environmental science | 4 068.24 | 1 166.25 | 5 234.49 | 0.6% | 22.3% | 23.0% |
| Art | 4 290.73 | 625.46 | 4 916.18 | 0.6% | 12.7% | 17.0% |
| History | 4 091.92 | 528.79 | 4 620.71 | 0.5% | 11.4% | 12.4% |
| Business | 2 967.66 | 712.52 | 3 680.18 | 0.4% | 19.4% | 18.1% |
| Geography | 2 848.64 | 811.28 | 3 659.92 | 0.4% | 22.2% | 21.3% |
| Sociology | 2 910.41 | 409.57 | 3 319.99 | 0.4% | 12.3% | 23.7% |
| Economics | 1 827.57 | 401.35 | 2 228.93 | 0.3% | 18.0% | 23.5% |
| **Total** | **582 273.00** | **279 447.00** | **861 720.00** | **100.0%** | **32.4%** | **21.4%** |

*4.2. Normalized Open Access Index*

The value of the Normalized Open Access Index (NOAI) for all disciplines combined is calculated by taking the weighted average of the NOAI values for each discipline. In this regard, the overall NOAI is established at 1.38. This value signifies that, on average, open access (OA) publications are 38% more prevalent in Non-Patent References (NPRs) than in the entire OpenAlex database. It suggests that NPRs have a higher propensity to include OA publications compared to the entire database, thus demonstrating a pronounced preference for open access to scientific literature among patent citations. The global NOAI value highlights the growing importance of open access in the research and innovation landscape, emphasizing the need to encourage and promote policies and practices that foster accessibility and dissemination of scientific knowledge.

Figure 1 presents the distribution of the Normalized Open Access Index (NOAI) across different disciplinary domains, focusing on publications with Digital Object Identifiers (DOIs). The NOAI values, ranging from 0.52 to 1.73, provide insights into the relative prevalence of Open Access (OA) publications within patent citations across various disciplines.

Several noteworthy trends emerge from the data. Disciplines such as Biology, Medicine, and Chemistry exhibit relatively high NOAI values, indicating a substantial reliance on open access literature within these fields. Conversely, disciplines like Sociology, Engineering, and Materials Science demonstrate lower NOAI values, suggesting a comparatively lower dependency on open access publications, even when considering those with DOIs. Interestingly, disciplines like Computer Science and Business present NOAI values that are higher than the average, despite their relatively lower shares of open access publications within the NPRs dataset.

In addition to the analysis presented, it would be intriguing to incorporate consideration of "black open access" (OA) sources, such as those accessible through platforms like Sci-Hub. Integrating black OA into the analysis could provide further insights into the accessibility and reliance on open access literature within different disciplines. Given the controversial nature of black OA and its potential impact on traditional publishing models, examining its influence on the Normalized Open Access Index (NOAI) could offer a more comprehensive understanding of scholarly communication practices.

Understanding how black OA sources contribute to the availability and utilization of research literature could reveal whether the NOAI values would be even higher than observed. This exploration could shed light on the extent to which applicants rely on alternative means to access scholarly content, particularly in disciplines where access barriers are more pronounced. Furthermore, it could prompt discussions on the efficacy of traditional publishing models and the role of black OA in democratizing access to knowledge.

**Figure 1 : Normalized Open Access Index by discipline (only publications with DOI)**

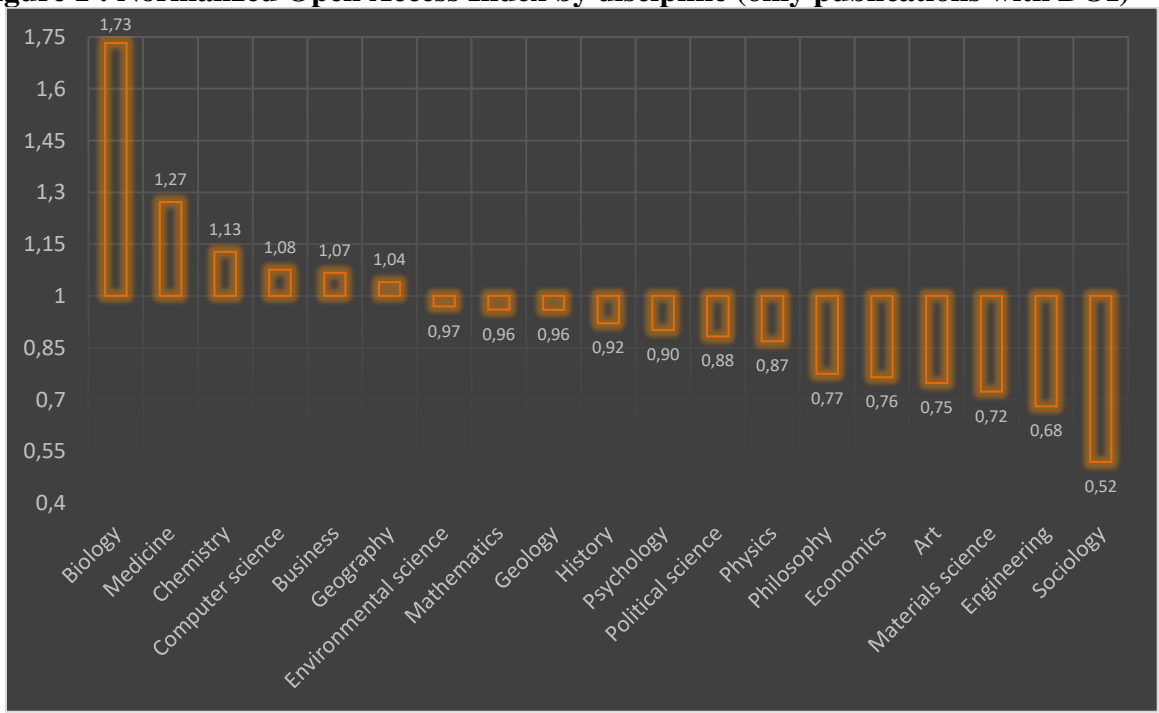

## 5. Discussion

The salient revelation of this study is that OA publications are more prevalent in patent citations compared to the overall OpenAlex database. This finding not only illuminates the symbiotic relationship between open science and inventive activity but also highlights the transformative potential of open access resources in driving technological innovation. It presents a novel insight into the utilization patterns of OA literature by inventors, reflecting a paradigm shift towards leveraging openly accessible scientific knowledge in the inventive process.

Our results are consistent with previous research. (Bryan & Ozcan, 2021) found a slight increase in citations to NIH-funded research following the NIH Open Access policy implementation, particularly benefiting small biotechnology companies. Similarly, (Jahn et al., 2022) discovered that approximately one third of patent families referenced non-patent literature, with a growing proportion of Open Access citations over time. These findings support the growing importance of Open Access in facilitating access to scientific knowledge and driving innovation.

The implications of these findings extend beyond academia, carrying significant political ramifications. The observed disparities in OA usage among different fields underscore the need for targeted policy interventions to promote open access awareness and enhance the visibility of OA resources. Policymakers can leverage these insights to formulate strategies aimed at fostering a more inclusive and collaborative research environment, ultimately driving societal progress and economic development.

Additionally, the pronounced preference for OA publications in patent citations signals a shift towards a more open and transparent innovation ecosystem. This presents an opportunity for policymakers to enact policies that incentivize the adoption of open science practices and promote equitable access to scientific knowledge. By fostering a culture of openness and collaboration, policymakers can stimulate innovation, accelerate scientific progress, and address societal challenges more effectively.

## 6. Conclusion

This study aimed to analyze the extent to which inventive activity relies on open science, particularly focusing on the utilization of Open Access (OA) publications within patent citations. Leveraging data from the (Marx, 2023) database, encompassing citations of patents to scientific publications (Non-Patent References - NPRs), and supplemented by OpenAlex data, our analysis provides valuable insights into the relationship between open science and innovation.

The finding that OA publications are more prevalent in patent citations than in the overall OpenAlex database underscores the significant impact of open science on inventive activity. Across all disciplines, OA publications are 38% more prevalent in patent citations (NPRs) than in the entire OpenAlex database, indicating a pronounced preference among inventors for utilizing openly accessible scientific literature in their innovative endeavors.

The substantial prevalence of OA publications, particularly evident in disciplines such as biology, medicine, chemistry, and computer science, reflects the increasing recognition of the value and relevance of open access resources in driving technological innovation. This trend highlights not only the growing accessibility of OA resources but also their perceived importance in shaping inventive processes and advancing scientific progress.

## 8. Competing interests
The author have no relevant financial or non-financial interests to disclose.